\documentclass[twocolumn,showpacs,preprintnumbers,amsmath,amssymb,aps]{revtex4}
\usepackage{bm}
\usepackage{latexsym}
\usepackage{eufrak}
\usepackage[latin1]{inputenc}
\usepackage{latexsym}
\usepackage{amsmath}
\usepackage{epsfig}
\usepackage{ifpdf}
\usepackage{graphicx}
\usepackage{dcolumn}
\usepackage{amsmath}
\usepackage{amsfonts}
\usepackage{amssymb}
\usepackage{color}

\begin{document}
\title{About the Geometric Solution to the Problems of Dark Energy.}
\author{Miguel A. Garc\'{\i}a-Aspeitia} 
\email{agarca@fis.cinvestav.mx}
\affiliation{Departamento de F\'{\i}sica, Centro de Investigaci\'on y de Estudios Avanzados del I.P.N.
Apdo. Post. 14-740 07000, D.F., M\'exico}

\affiliation{Escuela Superior de Ingenier\'ia Textil del IPN. Av. Instituto Polit\'ecnico Nacional S/N, Edificio  8, 
Unidad Profesional Adolfo L\'opez Mateos, Col. Lindavista. M\'exico, D.F., C.P. 07300. }

\begin{abstract}
In this paper is proposed a geometric solution to the dark energy, assuming that the space can be divided into regions of size $\sim L_{p}$ and energy $\sim E_{p}$. Significantly this assumption generate a energy density similar to the energy density observed for the vaccum energy, the correct solution for the coincidence problem and the state equation characteristic of quintessence in the comoving coordinates. Similarly is studied the ultraviolet and infrarred limits and the amount of dark energy in the Universe.
\end{abstract}

\draft
\date{\today}
\maketitle

\section{Introduction.}

One of the most intriguing problem of the modern cosmology is the accelerated expansion of the Universe. 
The best explanation for this problem, is the existence of an unknown kind of energy not predicted by the standard model of particles nor 
by the general theory of relativity. This dark component is called dark energy (DE) with the property of accelerate the rate expansion of the Universe. 
Currently exist differents kind of models trying to explain the nature and the behavior of DE. For example the best models for DE are quintessence, 
phantom energy, cosmological constant and 
higher dimensional theories, each one of them try to mimic the behavior of DE. However the discordance between the theoretical predictions and the observations 
puts into question the validity of the models at large and Planck scales.

To understand the detail of DE we will enumerate the main problems in the following way

\begin{enumerate}

\item \emph{The fine tuning problem.} Observational evidence show that the energy density today 
must be $\vert\rho^{obs}\vert\leqslant2\times10^{-10}erg/cm^{3}$ \cite{Carroll}, this imply that the theoretical 
predictions must relate two quantities appear to be unrelated \cite{Bousso1}

\begin{equation}
t_{DE}\sim t_{obs} \label{time}
\end{equation}
where $t_{DE}\sim\rho_{DE}^{-1/2}$ is the domination time of the DE and $t_{obs}$ is the time at which observer exist.

\item \emph{The coincidence problem.} Another problem caused by DE is the coincidence where this 
problem can be summarized in the following question: Why the Universe starts the acceleration today ($\sim13.7\times10^{9} yrs$)? 
Any good model should address this question. 

\item \emph{The "central" Problems.} At this point we refer to the main features of DE as: the amount of DE in 
the Universe, the state equation and the convergence of $\langle\rho\rangle$ in the infrarred and ultraviolet limit.

\end{enumerate}

In the following sections we will focuse on addressing one by one the above points with the aim of found some physical explanation to the problem of DE.

In the following, is used the CGS units, unless explicitly written.
 
\section{The three main problems of dark energy.}
 
\subsection{The fine tuning problem.}

Before to mention the proposal, it is important to stress that we work in a \emph{physical coordinates} \cite{Liddle}.

\emph{The proposal.} The idea is assume that the Universe is full of minimal regions of equal size \cite{Miguel}. It is assumed that a region smaller that the one given by the equation \eqref{Lp} will collapse into a Planckian black hole where all the physical information is lost. Then it is possible to write this minimal regions as

\begin{equation}
L_{p}=\frac{1}{\sqrt{2}}\sqrt{\frac{\hbar G}{c^3}}\approx\sqrt{\frac{\hbar G}{c^3}}, \label{Lp}
\end{equation}
where $G$ is the gravitational Newton constant, $c$ is the light velocity and $\hbar$ is the reduced Planck constant. It is important to remark that the last expression in the equation \eqref{Lp} is the well known Planckian longitude.

\begin{figure}[htp]
\centering
\includegraphics[scale=0.32]{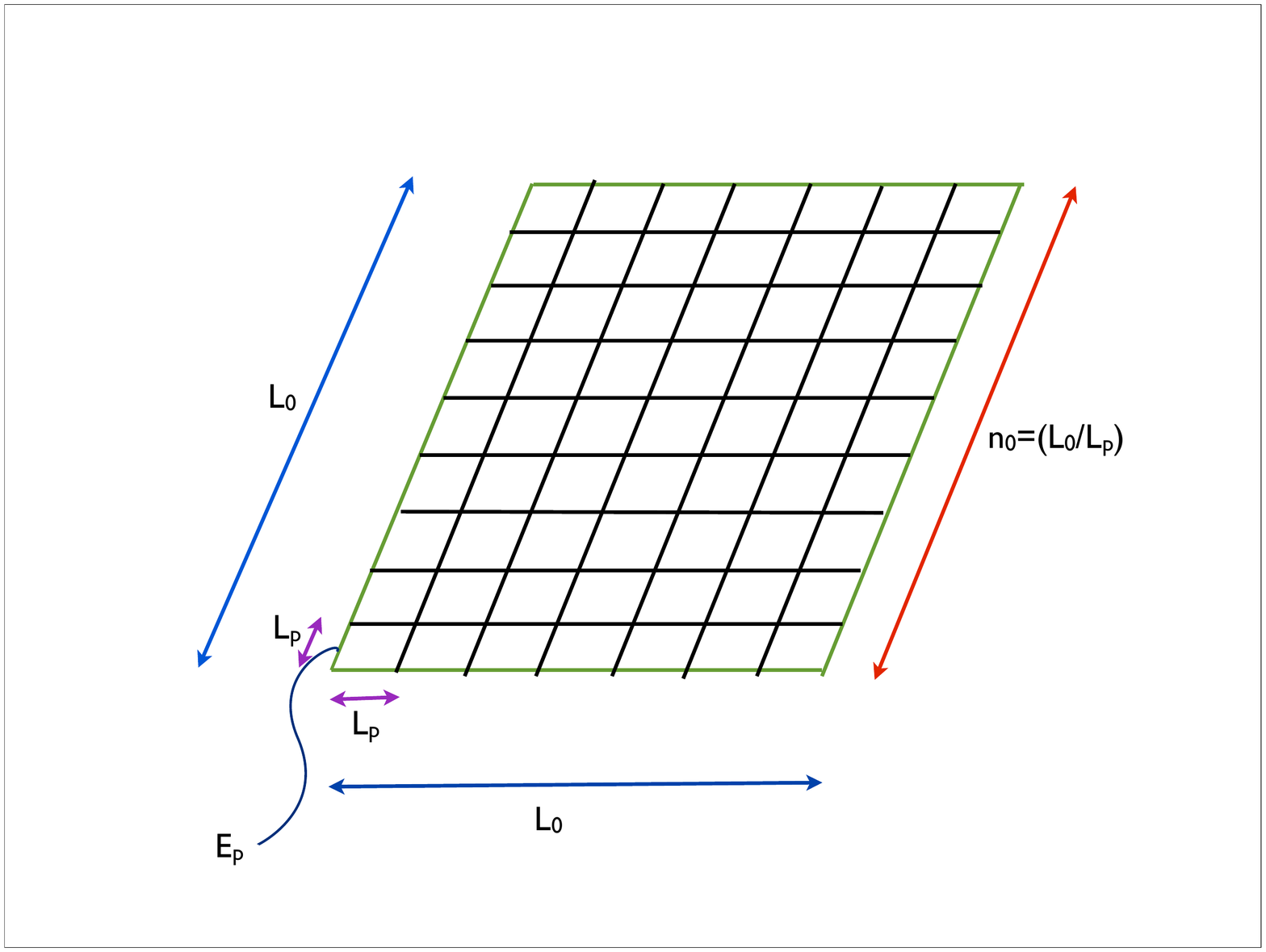}
\caption{Sketch of the grid hypothesis. In the figure it is observed the minimal regions of the "bricks" $L_{p}$ and the minimal energy $E_{p}$ as well as the size of the Universe $L_{0}$ and the preferential direction of the counting $n_{0}$.} \label{fig:1}
\end{figure}

Then, intuitively we assume that inside of compact region of the Universe with size $L_{0}$ exist $n_{0}$ minimal regions of size $L_{p}$ immersed in the following way
\[n_{0}=\left(\frac{L_{0}}{L_{p}}\right).\]
It is possible to observe that the we made a count of the number of Planck regions in a preferred direction and not in the three spatial directions of the space (see Figure \ref{fig:1}). The answer to why this is done can have deeper implications, which will be discussed in the section IV.

Returning to the idea, it is possible to assume that each region have a energy \cite{Miguel} written as

\begin{equation}
E_{p}=\frac{1}{2\sqrt{2}}\sqrt{\frac{\hbar c^{5}}{G}}\approx\sqrt{\frac{\hbar c^{5}}{G}}, \label{Ep}
\end{equation}
where the last expression of the equation \eqref{Ep} is the Planck energy. Then it is possible to obtain the total 
energy provided by all the regions in the Universe as
\[\langle E_{T_{0}}\rangle\approx\sum_{i=1}^{n_{0}}E_{p_{i}}=n_{0}E_{p}.\]
On the other hand, observational evidence show that the Universe at large scales behaves as homogeneous 
and isotropic space expanding in time with a flat geometry. Flatness implies that the geometry of the hypersurface 
is Euclidean $\mathbb{R}^3$ then, it is possible to define the volumen of the Universe as \cite{Bousso}
\[V_{0}\approx\frac{4}{3}\pi L^{3}_{0}.\]
Defining the energy density as $\langle\rho_{Y}\rangle=\langle E_{T_{0}}\rangle/V_{0}$ it is possible to write in the following way

\begin{equation}
\langle\rho_{Y}\rangle_{L_{0}}\approx\frac{3}{4\pi L_{0}^{3}}\sum_{i=1}^{L_{0}/L_{p}}E_{pi}\approx\left(\frac{3c^4}{8\pi G}\right)L_{0}^{-2}. \label{enrgdens}
\end{equation}
Now, using the evidence that the Universe is finite it is possible to compactified 
in a hypersurface and assign an approximate size of the Universe actually as $L_{0}\approx ct_{0}$, at a fixed moment of time, 
where $t_{0}$ is the actual age of the Universe \cite{Holo}. Then the previous expression \eqref{enrgdens} can be written as

\begin{equation}
\langle\rho_{Y}\rangle_{t_{0}}\approx\left(\frac{3c^2}{8\pi G}\right)t_{0}^{-2}, \label{t}
\end{equation}
where it is shown the comparison between the present cosmological time and the energy density of the DE as $t_{0}\sim\rho_{Y}^{-1/2}$. Helping to solve the 
fine tuning problem \eqref{time} where $\langle\rho_{Y}\rangle_{t_{0}}\approx8.62\times10^{-9}erg/cm^{3}$.

\subsection{The Recent Acceleration.}

Another problem for the dark energy is the coincidence of the recent acceleration ($\sim4.32\times10^{17} s$).  
In this model, we explore a possible solution to the problem in the following way:

In fact, the Universe has a specific quantity of barions, radiation, neutrinos and dark matter with the 
possibility of colapse the Universe with its gravitational interactions. 
Then, it is possible to obtain with observations the total mass of the Universe as $M_{u}\sim10^{56}g$. With this amount 
of matter it is necessary a minimum of energy to accelerate the Universe which can be written as
\[E_{min}=n_{acc}E_{p}\gtrsim G\frac{M_{u}^2}{L_{acc}},\]
where $n_{acc}=(L_{acc}/L_{p})$. Assuming that we know the Newtonian potential, it is straightforward to demonstrate the following equation

\begin{equation}
L_{acc}\gtrsim\sqrt{\frac{GL_{p}}{E_{p}}}M_{u}=\frac{\sqrt{2}GM_{u}}{c^{2}}\approx R_{s}, \label{accel}
\end{equation}
where $L_{acc}$ is the size of the Universe in the moment of acceleration. It is possible to observe \eqref{accel} that the minimal longitude must be approximately the Schwarzschild radius $R_{s}$.

Adding numbers to the last equation is obtained $L_{acc}\gtrsim1.0482\times10^{28}$ $cm$ $\approx L_{0}$ this imply that $t_{acc}\approx3.49\times10^{17} s$ which coincides in a good way with the moment of acceleration. 

The Universe start the acceleration at this age and never before while the mass is of $M_{u}\sim10^{56}g$.

\subsection{The Amount of Dark Energy in the Universe.}

In cosmology, the critical energy density relate the content in the Universe with its geometry and is defined as

\begin{equation}
\rho_{crit}(t_{0})=\frac{3H^2_{0}c^{2}}{8\pi G}, \label{crit}
\end{equation}
where $H_{0}$ is the Hubble rate today. If $\rho>\rho_{crit}$ the geometry is $\mathbb{S}^{3}$, $\rho\sim\rho_{crit}$ the geometry is $\mathbb{R}^{3}$ and $\rho<\rho_{crit}$ the geometry is $\mathbb{H}^{3}$.

On the other hand, the equation \eqref{t} can be written in terms of the Hubble parameter $H(t_{0})=H_{0}=t_{0}^{-1}$ due to the dimensionality between the Hubble parameter and the time $[H_{0}]=[seg^{-1}]=[t_{0}^{-1}]$. Therefore, it is possible to write

\begin{equation}
\langle\rho_{Y}\rangle_{t_{0}}\approx\frac{3H^{2}_{0}}{8\pi G}c^2, \label{DE}
\end{equation}
using dimensional analysis. Defining the density parameter as $\Omega_{Y}\equiv\langle\rho_{Y}\rangle/\rho_{crit}$ it is possible to obtain
\[\Omega_{Y}\approx1.\]
For this reason we assume that this component is the dominant component in the actual Universe, coinciding with the observations about the dark energy which is more than $70\%$ of the known Universe.

\section{The Limits and the State Equation.}

\subsection{The Ultraviolet and Infrarred Limit.}

Now, the questions is: What is the two limits of the equation \eqref{enrgdens}? What is the behavior of the energy density in the singularity? and What is the behavior of the energy density in infinity?

\begin{enumerate}

\item \emph{Ultraviolet Limit ($L\to\infty$).} When the Universe tends to infinity, the equation \eqref{enrgdens} can be written as

\begin{equation}
\lim_{L\to\infty}\langle\rho_{Y}\rangle_{L}\approx\lim_{L\to\infty}\frac{3}{4\pi L^{3}}\sum_{i=1}^{L/L_{p}}E_{pi}\to0, \label{limsup}
\end{equation}
this imply that the equation of the energy density for the DE is convergent to zero in infinity reducing the energy density as the Universe grows. 

In comparison, the integral energy density is divergent in infinity, for this reason must be calculated with a wave number cutoff \cite{Weinberg} as
\[\langle\rho\rangle=\int_{0}^{\Lambda}\frac{4\pi k^{2}c^{2}dk}{(2\pi\hbar)^{3}}\frac{1}{2}\sqrt{k^{2}+m^{2}}\approx\frac{\Lambda^{4}}{16\pi^{2}},\]
where $\Lambda>>m$.

\item \emph{Infrarred Limit ($L\to L_{p}$).} On the other when the Universe decreases it is assumed that the minimal length will be $L_{p}$, then from the equation \eqref{enrgdens} it is obtained the following expression

\begin{eqnarray}
\lim_{L\to L_{p}}\langle\rho_{Y}\rangle_{L}&&\approx\lim_{L\to L_{p}}\frac{3}{4\pi L^{3}}\sum_{i=1}^{L/L_{p}}E_{pi}\nonumber\\&&\approx\frac{3c^{7}}{4\pi G^{2}\hbar}, \label{liminf}
\end{eqnarray}
the overwhelming Planck energy density shown in the equation \eqref{liminf} could plunge us into a Universe always dominated by the dark sector, however inflation could help us to solve the problem because the exponential growth of the e-foldings permit the entry of a greater amount of space controlling the growth of the energy provided by the Planck regions (see equation \eqref{limsup}).

\end{enumerate}

\subsection{The State Equation.}

Now we focuse in obtain the state equation for this characteristic density $\rho_{Y}$. Then from the equation \eqref{enrgdens} it is possible to write

\begin{equation}
\langle\rho_{Y}\rangle\approx\rho_{0}a^{-2}, \label{III1}
\end{equation}
where \[\rho_{0}=\frac{3c^{2}}{8\pi Gd^{2}},\]
where we fix at a comoving coordinates system $\vec{L}=a(t)\vec{d}$ and $a(t)$ is the scale factor \cite{Liddle}. On the other hand the state equation can be written as $\omega_{Y}\equiv\langle p_{Y}\rangle/\langle\rho_{Y}\rangle$ and the conservation law in a comoving system reads as

\begin{equation}
\frac{\partial}{\partial t}\langle\rho_{Y}\rangle+3\frac{\dot{a}}{a}\left[\langle\rho_{Y}\rangle+\langle p_{Y}\rangle\right]=0, \label{III3}
\end{equation}
with the equations \eqref{III1} and \eqref{III3} the pressure can be written as $\langle p_{Y}\rangle\approx-(1/3)\rho_{0}a^{-2}$, this implies that the state equation reads as
\[\omega_{Y}\approx-\frac{1}{3},\]
where the last equation is clearly similar to the state equation of quintessence $-1<\omega<-1/3$ where the energy density decreases with the scale factor as $\rho_{Q}\varpropto a(t)^{-3(1+\omega)}$.

\section{Discussion.}

It is important to remark that the proposed hypothesis generates a good results to the problems of dark energy. This may mean that the Universe is made of minimum regions $\sim L_{p}$ and minimum energy $\sim E_{p}$ such as a quantum grid.

An important feature is the count in a preferential direction of the number of regions $n_{0}$ in the grid, this has 
important consequences in the understand of the microscopic structure of the space time. For example, if we do not consider a preferential direction and do the count in the three spatial directions $n_{0}^{3}$ (just as intuition tell us) we obtain the following expression for the energy density
\[\langle\rho^{3D}\rangle\approx\frac{3}{4\pi}\left(\frac{E_{p}}{L^{3}_{p}}\right)=\rho_{0}^{3D}=cte,\]
where the numerical value of the last expression is $\langle\rho^{3D}\rangle\approx1.3866\times10^{113}erg/cm^{3}$ regaining the problem of vaccum energy implemented by the energy density integral calculated by \citet{Weinberg}. Even more, using the state equation and the conservation law \eqref{III3} it is possible to obtain
\[\langle p^{3D}\rangle\approx-\langle\rho^{3D}\rangle,\]
where the last equation is associated with the cosmological constant $\omega\approx-1$.

Summarizing, if we perform the count in a preferred direction, we get the values needed to adjust to the observations of dark energy being quintessence the model associated with that hypothesis. On the other hand if we perform the count in the three dimensional directions we regaining the problem of the vaccum energy and the associated model is the cosmological constant.

This leads us to believe that there is a preferential direction which the counting is done, violating Lorentz symmetry. However an interesting alternative \emph{ad hoc} to this model (instead of considering a preferential direction) is consider that the Universe can be one dimensional at short scales and at large scales behave as a four dimensional Universe as the proposition of \citet{Dimensions} avoiding the preferential direction.

It should be noted that this model can be considered as a toy model, where the results should be formalized, however can help us to clarify questions about dark energy currently unsolved. Further investigation may be made to study inflation since not only could help us to the problem of cosmic acceleration, but it also could provide clues about inflation.

\section{Acknowledgement.}

I thank Yasm\'in Alc\'antara for encouragement to publish this result. In the same way I would like to acknowledge the help and the enlightening conversation of Abdel P\'erez-Lorenzana and Juan Maga\~na. This work was partially supported by CONACyT M\'exico, under grant 49865-F, 216536/219673, Instituto Avanzado de Cosmologia (IAC) collaboration.

\end{document}